\documentclass[conference,onecolumn,letterpaper,final]{IEEEtran}
\usepackage{caption}
\usepackage{subcaption}  
\usepackage{algorithm}
\usepackage{algpseudocode}
\usepackage{algorithm} 
\usepackage{longtable}

\usepackage{titlesec}

\usepackage[T1]{fontenc}
\usepackage{lmodern} 
\usepackage{amssymb,stmaryrd,amsmath,amsfonts,rotating}
\usepackage{amsmath,amsthm,amssymb,cases}
\usepackage{tikz}
\usetikzlibrary{arrows}
\usetikzlibrary{tikzmark}
\usetikzlibrary{matrix}
\usetikzlibrary{positioning}
\usepackage{algpseudocode}
\usepackage{amsmath,cases}
\usepackage{bbm}
\usepackage{amsthm}
\usepackage{booktabs}
\usepackage{color}
\usepackage{amssymb}
\usepackage{bm}
\usepackage{thmtools}

\theoremstyle{definition}
\declaretheoremstyle[style=definition,qed=\openbox,]{ppstyle}

\usepackage{mathtools}
\mathtoolsset{showonlyrefs}

\usepackage{mathtools}
\usepackage{amsmath}

\usepackage{bbm}
\usepackage{mathrsfs}

\makeatletter
\newcommand\RedeclareMathOperator{%
  \@ifstar{\def\rmo@s{m}\rmo@redeclare}{\def\rmo@s{o}\rmo@redeclare}%
}
\newcommand\rmo@redeclare[2]{%
  \begingroup \escapechar\m@ne\xdef\@gtempa{{\string#1}}\endgroup
  \expandafter\@ifundefined\@gtempa
     {\@latex@error{\noexpand#1undefined}\@ehc}%
     \relax
  \expandafter\rmo@declmathop\rmo@s{#1}{#2}}
\newcommand\rmo@declmathop[3]{%
  \DeclareRobustCommand{#2}{\qopname\newmcodes@#1{#3}}%
}
\@onlypreamble\RedeclareMathOperator
\makeatother

\definecolor{grey}{rgb}{0.7, 0.75, 0.71}

\def\dark-red#1{\textcolor[rgb]{0.7,0.0,0.0}{#1}}

\definecolor{amber}{rgb}{1.0, 0.75, 0.0}

\def\...{\dotsc}

\def\intT2{\int_{-T/2}^{T/2}}

\def\sumi1n{\sum_{i=1}^{n}}
\def\sumi1N{\sum_{i=1}^{N}}
\def\sumi0N--{\sum_{i=0}^{N-1}}

\def\=def{\overset{\text{\small def}}{=}}

\DeclarePairedDelimiterX{\inp}[2]{\langle}{\rangle}{#1, #2}

\def\<{\langle}
\def\>{\rangle}

 \def\mat4#1#2#3#4{
\begin{pmatrix}
 #1&\ccc&#2\\
 \vdots&&\vdots\\
 #3&\ccc&#4
\end{pmatrix}}

\def\0sf{\mathsf{0}}
\def\1sf{\mathsf{1}}

\def\0BS{\boldsymbol{0}}
\def\1BS{\boldsymbol{1}}

\def\0B{\mathbf{0}}
\def\1B{\mathbf{1}}

\def\0H{\hat{0}}
\def\1H{\hat{1}}

\def\+TT{\texttt{+}}
\def\-{\texttt{-}}
\def\+KB{|+\> \<+|}
\def\-KB{|-\> \<-|}

\def\q0{|0\>}

\def\0U{\underline{0}}
\def\1U{\underline{1}}

\def\0UH{\underline{\0H}}
\def\1UH{\underline{\1H}}

\RedeclareMathOperator{\Im}{Im}

\ifCLASSINFOpdf
\else
\fi
\begin{document}
%
\title{Random Construction of Quantum LDPC Codes}
\author{
\IEEEauthorblockN{Koki Okada and Kenta Kasai}
\IEEEauthorblockA{
Institute of Science Tokyo\\
Email: okada.k.3154@m.isct.ac.jp, kenta@ict.eng.isct.ac.jp}
}
\maketitle

\begin{abstract}
We propose a method for modifying orthogonal sparse matrix pairs used in CSS codes while preserving their matrix row and column weight distributions, which play a crucial role in determining the performance of belief-propagation decoding. 
Unlike simple row or column permutations that merely reorder existing elements, the proposed local modification introduces genuine structural randomness through small $2\times2$ cross-swap operations followed by integer-linear-program-based local repairs that restore orthogonality. 
By applying this procedure repeatedly in a random manner, ensembles of randomized quantum LDPC codes can be constructed. 
The computational complexity of each repair depends only on the maximum row and column weights and is independent of the overall matrix size, ensuring scalability to large code blocks.
\end{abstract}
\IEEEpeerreviewmaketitle

\section{Introduction}

Quantum low-density parity-check (LDPC) codes have emerged as promising candidates for scalable and fault-tolerant quantum computation due to their low decoding complexity~\cite{mackay2004sparse,komoto2024quantumerrorcorrectionnear}.
Among them, CSS-type quantum LDPC codes, defined by two binary parity-check matrices \(H_X\) and \(H_Z\) satisfying \(H_X H_Z^{\mathrm{T}} = 0\), provide a flexible framework for constructing large quantum codes from classical LDPC components.

For classical LDPC codes, various construction techniques have been established, such as Gallager's random construction~\cite{Gallager1962}, protograph-based designs~\cite{Thorpe2003}, and the random graph approach~\cite{Richardson:2008:MCT:1795974}.
Since the decoding threshold of belief-propagation (BP) decoding is primarily determined by the weight distributions of parity-check matrices, it is desirable to randomize matrix connectivity while preserving these distributions.
These classical constructions are designed so that the row and column weight distributions can be explicitly specified.
Moreover, it is known that classical LDPC codes with column weight three or higher typically achieve a minimum distance that grows linearly with the code length \cite{Richardson:2008:MCT:1795974}.
Therefore, in the classical setting, the minimum-distance design has not been a major concern.

In contrast, for quantum LDPC codes, constructions that exploit cyclic structures—such as bicycle codes~\cite{mackay2004sparse} and quasi-cyclic (QC) LDPC codes~\cite{4557323}—allow relatively flexible control over the degree distributions.
However, since these designs typically rely on generating a large orthogonal matrix pair and then removing redundant rows to form a valid code, the resulting minimum distance tends to be small.
On the other hand, tensor-product-based constructions such as hypergraph-product (HGP) codes~\cite{TillichZemor2014} can be designed to achieve large minimum distances, but it is difficult to control their degree distributions precisely.
Therefore, it is desirable to develop highly random construction methods that do not impose strong structural constraints, in analogy to the random ensemble constructions of classical LDPC codes.

The main idea of our approach is as follows.
We first prepare an initial pair of orthogonal matrices \( (H_X, H_Z) \) having the desired degree distributions.
For instance, a pair of regular matrices with column weight \( d_c \) and row weight \( d_r \) can be easily constructed by tiling identity matrices in a block array, which trivially satisfies \( H_X H_Z^{\mathrm{T}} = 0 \).
Next, we introduce a small random modification to \( H_X \) that preserves its degree distribution.
This perturbation typically breaks the orthogonality condition, so we locally repair \( H_Z \) by adjusting a small number of its entries while keeping its row and column weights unchanged.
The repair process is formulated as a compact integer linear program that enforces both orthogonality and weight preservation.
By repeating this procedure randomly, we can generate an ensemble of orthogonal sparse matrix pairs that maintain the same degree distributions but exhibit distinct random connectivity patterns.

The proposed random modification framework enables the systematic generation of quantum LDPC codes that preserve the same degree distributions as the original structured designs while introducing genuine randomness into their connectivity.
Because the asymptotic decoding behavior is primarily governed by these weight distributions, the proposed method offers a way to explore randomized ensembles consistent with desired structural properties.
In this work, we focus on the design methodology itself and the evaluation of its computational complexity, leaving the analysis of belief-propagation decoding performance for future studies.

The computational cost of each local repair depends only on the maximum row and column weights and is independent of the overall matrix size, which ensures scalability to large block lengths.
This property makes the proposed method suitable for constructing large and potentially high-performance quantum LDPC codes with controllable degree distributions.

The remainder of this paper is organized as follows.
Section~II reviews the orthogonality condition and the structure of the initial matrix pair.
Section~III introduces the random modification operation and the formulation of the local repair problem.
Section~IV presents the integer-linear-program-based repair algorithm and discusses its computational complexity.
Finally, Section~V concludes the paper with remarks on scalability and future directions.

\section{Local Random Modification and Repair}

This section describes a random modification framework for generating ensembles of orthogonal sparse matrix pairs $(H_X,H_Z)$ with arbitrary prescribed weight distributions.
Unlike simple row or column permutations, which merely reorder existing entries, the proposed approach introduces genuine structural randomness by locally exchanging elements within $2\times2$ submatrices.
Starting from an initial pair of binary matrices satisfying the orthogonality condition $H_X H_Z^{\mathsf T}=0$, 
we apply small random perturbations—called cross swaps—to $H_X$ and then locally repair $H_Z$ to restore orthogonality.
The repair process is formulated as a compact integer linear program (ILP) that simultaneously enforces both orthogonality and the preservation of row and column weight distributions.
By iteratively applying these local modifications and repairs, we can construct large ensembles of well-randomized orthogonal matrix pairs that are structurally distinct, while maintaining the same degree profiles relevant to decoding performance.
\subsection{Initial Structure and Orthogonality}

Let $H_X$ and $H_Z$ be binary parity-check matrices over  that satisfy the orthogonality condition
\[
H_X H_Z^{\mathsf T} = 0.
\]
Each matrix may have an arbitrary prescribed row and column weight distribution.
Such a pair $(H_X,H_Z)$ serves as the initial orthogonal structure for our random modification procedure.

For example, a regular pair can be constructed by tiling the $P\times P$ identity matrix $I_P$ in a $d_c\times d_r$ block array,
\[
H_X = H_Z =
\begin{bmatrix}
I_P & I_P & \cdots & I_P\\
\vdots & \vdots & \ddots & \vdots\\
I_P & I_P & \cdots & I_P
\end{bmatrix},
\]
which satisfies $H_X H_Z^{\mathsf T}=0$ and has constant row and column weights $d_r$ and $d_c$, respectively.
In general, however, we allow $H_X$ and $H_Z$ to have arbitrary weight distributions, provided that orthogonality is maintained.
Preserving these distributions throughout the random modification process is essential for maintaining decoding performance.

\subsection{Cross Swap Operation}
Given a binary matrix $H$, choose row indices $(i_1,i_2)$ and column indices $(j_1,j_2)$ such that
\[
H_{i_1,j_1}=H_{i_2,j_2}=1,\quad
H_{i_1,j_2}=H_{i_2,j_1}=0.
\]
Replacing the submatrix
\(
\begin{bmatrix}
1 & 0\\
0 & 1
\end{bmatrix}
\)
with
\(
\begin{bmatrix}
0 & 1\\
1 & 0
\end{bmatrix}
\)
is called a \textit{cross swap} (also referred to as a $2\times2$ switch).
This operation exchanges the diagonal elements of a $2\times2$ submatrix while preserving the row and column weights.

\subsection{Localization of the Violation}
Let the effect of a cross swap on orthogonality be localized by the following index sets:
\[
I=\{i\mid (H_X'H_Z^{\mathsf T})_{*,i}\neq0\},\quad
J=\{j\mid (H_Z)_{I,j}\neq0\},\quad
K=\{k\mid (H_X')_{k,J}\neq0\}.
\]
Then the orthogonality condition can be written as
\[
(H_X')_{K,J}(\Delta_{I,J})^{\mathsf T}=(H_X'H_Z^{\mathsf T})_{K,I},
\]
which yields a system of GF(2) linear equations in the unknown repair matrix $\Delta_{I,J}$.

\subsection{Weight-Preservation Constraints}
We introduce the ``signed balance'' constraints
\[
\sum_{j\in J}\Delta_{i,j}(1-2H_{Z,i,j})=0,\quad
\sum_{i\in I}\Delta_{i,j}(1-2H_{Z,i,j})=0.
\]
These enforce that the number of $1\to0$ and $0\to1$ flips is the same in each affected row/column, hence the row/column weights of $H_Z'$ match those of $H_Z$ exactly.

\subsection{ILP-Based Local Repair Formulation}
Let $\Delta_{i,j}\in\{0,1\}$ be decision variables.
We seek a repair that satisfies
\begin{align*}
A_{\mathrm{big}} \operatorname{vec}(\Delta_{I,J}) &\equiv b_{\mathrm{big}} \pmod{2},\\
\sum_{j \in J} \Delta_{i,j} (1 - 2 H_{Z,i,j}) &= 0,\quad i \in I,\\
\sum_{i \in I} \Delta_{i,j} (1 - 2 H_{Z,i,j}) &= 0,\quad j \in J.
\end{align*}
Here $A_{\mathrm{big}}$ and $b_{\mathrm{big}}$ represent the GF(2) linear system induced by the orthogonality requirement $H_X H_Z^{\mathsf T}=0$.
The first line encodes orthogonality; the latter two lines enforce preservation of the row and column weights (the number of ones).

To cast the GF(2) system as an ILP, introduce $x=\operatorname{vec}(\Delta_{I,J})\in\{0,1\}^{v}$ and slack variables $s\in\mathbb{Z}_{\ge0}^{m}$, and rewrite the parity constraints $A_{\mathrm{big}}x\equiv b_{\mathrm{big}}\pmod2$ as
\[
A_{\mathrm{big}} x - 2s = b_{\mathrm{big}}.
\]
Here $v=|I|\,|J|$ is the number of variables and $m=|K|\,|I|$ is the number of orthogonality equations.
This guarantees equality modulo 2 via integrality.
As an objective, we minimize the total number of flips, $\sum x_{ij}$, to obtain the smallest repair that restores both orthogonality and the weight distributions.

\subsection{Algorithm}
\begin{enumerate}
  \item Initialize an orthogonal pair $(H_X,H_Z)$.
  \item Apply a random $2\times2$ cross swap; when orthogonality is violated, extract the affected index sets $I,J,K$.
  \item Formulate the above constraints as an ILP and solve the integer optimization with the OR-Tools CP-SAT solver.
  \item Set $H_Z' = H_Z \oplus \Delta$ and verify orthogonality and weight preservation.
  \item If successful, use $(H_X',H_Z')$ as the initial pair for the next iteration.
 \item In the subsequent iteration, exchange the roles of $H_X$ and $H_Z$ and repeat Steps~2--5.
\end{enumerate}

\section{Computational Complexity Evaluation}

In this section, we analyze the computational cost of a single local repair step in the proposed random modification algorithm.
Let $d_c$ and $d_r$ denote the maximum column and row weights of the parity-check matrices, respectively.
Each random $2\times2$ switch affects at most two columns of $H_X$, and thus up to $|I|\le 2d_c$ rows of $H_Z$ may lose orthogonality.
Since each affected row connects to at most $d_r$ columns, the number of affected columns is bounded by $|J|\le 2d_c d_r$.

The repair problem is defined on the submatrix $\Delta_{I,J}$ with $v=|I||J|$ binary decision variables.
Hence, the number of variables is bounded by
\[
v \le 4\, d_c^2 d_r .
\]
The orthogonality constraints in the ILP correspond to the parity equations induced by $H_X'H_Z^{\mathrm{T}}=0$.
Each affected column is connected to at most $d_c$ rows of $H_X'$, giving at most $|K|\le |J|d_c\le 2d_c^2 d_r$ parity-check equations.
Consequently, the total number of parity constraints is bounded by
\[
m = |K||I| \le 4\, d_c^3 d_r .
\]
In addition, two sets of signed-balance constraints are imposed to preserve the row and column weight distributions:
\[
|I| + |J| = O(d_c d_r).
\]
These equations are linear in $\Delta_{I,J}$ and contribute negligibly to the total complexity for small $d_c$ and $d_r$.

The memory footprint of each ILP instance scales as $O(v+m)=O(d_c^3 d_r)$, which depends only on the local degrees and not on the overall matrix size.
The runtime of the CP-SAT solver grows approximately linearly with $v+m$ in practice, because each subproblem is compact and independent.
For typical LDPC parameters such as $(d_c,d_r)=(3,8)$, the number of variables in each local ILP remains on the order of a few hundred (about $v\!\approx\!288$) with fewer than one thousand parity constraints, and each repair is solved within a few seconds on a standard laptop.

Overall, the computational complexity of the proposed randomization framework is dominated by the local ILP solving step, whose cost depends only on the maximum row and column weights and does not grow with the code length.
As in conventional LDPC code design, this property guarantees scalability to extremely large block lengths while exactly preserving both orthogonality and the intended degree distributions.

\section{Example}
\medskip
\noindent\textbf{Initial pair.}
We start from an orthogonal regular pair $(H_X,H_Z)$ obtained by tiling $P\times P$ identity matrices in a $d_c\times d_r$ array, which preserves row/column weights and satisfies $H_X H_Z^{\rm T}=0$. 
In the following toy example we apply one random $2\times2$ switch to $H_X$.
\[
\setlength{\arraycolsep}{1pt}
H_X=H_Z
=\begin{bmatrix}
1&0&0&0&1&0&0&0&1&0&0&0&1&0&0&0&1&0&0&0&1&0&0&0&1&0&0&0&1&0&0&0\\
0&1&0&0&0&1&0&0&0&1&0&0&0&1&0&0&0&1&0&0&0&1&0&0&0&1&0&0&0&1&0&0\\
0&0&1&0&0&0&1&0&0&0&1&0&0&0&1&0&0&0&1&0&0&0&1&0&0&0&1&0&0&0&1&0\\
0&0&0&1&0&0&0&1&0&0&0&1&0&0&0&1&0&0&0&1&0&0&0&1&0&0&0&1&0&0&0&1\\
1&0&0&0&1&0&0&0&1&0&0&0&1&0&0&0&1&0&0&0&1&0&0&0&1&0&0&0&1&0&0&0\\
0&1&0&0&0&1&0&0&0&1&0&0&0&1&0&0&0&1&0&0&0&1&0&0&0&1&0&0&0&1&0&0\\
0&0&1&0&0&0&1&0&0&0&1&0&0&0&1&0&0&0&1&0&0&0&1&0&0&0&1&0&0&0&1&0\\
0&0&0&1&0&0&0&1&0&0&0&1&0&0&0&1&0&0&0&1&0&0&0&1&0&0&0&1&0&0&0&1\\
1&0&0&0&1&0&0&0&1&0&0&0&1&0&0&0&1&0&0&0&1&0&0&0&1&0&0&0&1&0&0&0\\
0&1&0&0&0&1&0&0&0&1&0&0&0&1&0&0&0&1&0&0&0&1&0&0&0&1&0&0&0&1&0&0\\
0&0&1&0&0&0&1&0&0&0&1&0&0&0&1&0&0&0&1&0&0&0&1&0&0&0&1&0&0&0&1&0&\\
0&0&0&1&0&0&0&1&0&0&0&1&0&0&0&1&0&0&0&1&0&0&0&1&0&0&0&1&0&0&0&1
\end{bmatrix}
\]
\medskip
\noindent\textbf{Random switch on $H_X$.}
We choose $(i_1,j_1)=(10,6)$ and $(i_2,j_2)=(7,23)$ and apply a $2\times2$ cross-swap, obtaining $H_X'$ (red entries mark the swapped submatrix).
\[
\setlength{\arraycolsep}{1pt}
H_X'=\begin{bmatrix}
1&0&0&0&1&0&0&0&1&0&0&0&1&0&0&0&1&0&0&0&1&0&0&0&1&0&0&0&1&0&0&0\\
0&1&0&0&0&1&0&0&0&1&0&0&0&1&0&0&0&1&0&0&0&1&0&0&0&1&0&0&0&1&0&0\\
0&0&1&0&0&0&1&0&0&0&1&0&0&0&1&0&0&0&1&0&0&0&1&0&0&0&1&0&0&0&1&0\\
0&0&0&1&0&0&0&1&0&0&0&1&0&0&0&1&0&0&0&1&0&0&0&1&0&0&0&1&0&0&0&1\\
1&0&0&0&1&0&0&0&1&0&0&0&1&0&0&0&1&0&0&0&1&0&0&0&1&0&0&0&1&0&0&0\\
0&1&0&0&0&1&0&0&0&1&0&0&0&1&0&0&0&1&0&0&0&1&0&0&0&1&0&0&0&1&0&0\\
0&0&1&0&0&0&1&0&0&0&1&0&0&0&1&0&0&0&1&0&0&0&1&0&0&0&1&0&0&0&1&0\\
0&0&0&1&0&0&\textcolor{red}{\mathbf{1}}&1&0&0&0&1&0&0&0&1&0&0&0&1&0&0&0&\textcolor{red}{\mathbf{0}}&0&0&0&1&0&0&0&1\\
1&0&0&0&1&0&0&0&1&0&0&0&1&0&0&0&1&0&0&0&1&0&0&0&1&0&0&0&1&0&0&0\\
0&1&0&0&0&1&0&0&0&1&0&0&0&1&0&0&0&1&0&0&0&1&0&0&0&1&0&0&0&1&0&0\\
0&0&1&0&0&0&\textcolor{red}{\mathbf{0}}&0&0&0&1&0&0&0&1&0&0&0&1&0&0&0&1&\textcolor{red}{\mathbf{1}}&0&0&1&0&0&0&1&0&\\
0&0&0&1&0&0&0&1&0&0&0&1&0&0&0&1&0&0&0&1&0&0&0&1&0&0&0&1&0&0&0&1
\end{bmatrix}
\]

\medskip
\noindent\textbf{Localization.}
The violation of orthogonality is confined to the index sets
\[
I=\{2,3,6,7,10,11\},\qquad
J=\{2,3,6,7,10,11,14,15,18,19,22,23,26,27,30,31\},\qquad
K=\{2,3,6,7,10,11\},
\]
as extracted from $H_X'$ and $H_Z$. 
In $H_X'$, the rows and columns in $K$ and $J$ are highlighted in red, and in $H_Z$, those in $I$ and $J$ are highlighted.
\[
\setlength{\arraycolsep}{1pt}
H_Z=\begin{bmatrix}
1&0&0&0&1&0&0&0&1&0&0&0&1&0&0&0&1&0&0&0&1&0&0&0&1&0&0&0&1&0&0&0\\
0&1&0&0&0&1&0&0&0&1&0&0&0&1&0&0&0&1&0&0&0&1&0&0&0&1&0&0&0&1&0&0\\
0&0&\textcolor{red}{1}&\textcolor{red}{0}&0&0&\textcolor{red}{1}&\textcolor{red}{0}&0&0&\textcolor{red}{1}&\textcolor{red}{0}&0&0&\textcolor{red}{1}&\textcolor{red}{0}&0&0&\textcolor{red}{1}&\textcolor{red}{0}&0&0&\textcolor{red}{1}&\textcolor{red}{0}&0&0&\textcolor{red}{1}&\textcolor{red}{0}&0&0&\textcolor{red}{1}&\textcolor{red}{0}\\
0&0&\textcolor{red}{0}&\textcolor{red}{1}&0&0&\textcolor{red}{0}&\textcolor{red}{1}&0&0&\textcolor{red}{0}&\textcolor{red}{1}&0&0&\textcolor{red}{0}&\textcolor{red}{1}&0&0&\textcolor{red}{0}&\textcolor{red}{1}&0&0&\textcolor{red}{0}&\textcolor{red}{1}&0&0&\textcolor{red}{0}&\textcolor{red}{1}&0&0&\textcolor{red}{0}&\textcolor{red}{1}\\
1&0&0&0&1&0&0&0&1&0&0&0&1&0&0&0&1&0&0&0&1&0&0&0&1&0&0&0&1&0&0&0\\
0&1&0&0&0&1&0&0&0&1&0&0&0&1&0&0&0&1&0&0&0&1&0&0&0&1&0&0&0&1&0&0\\
0&0&\textcolor{red}{1}&\textcolor{red}{0}&0&0&\textcolor{red}{1}&\textcolor{red}{0}&0&0&\textcolor{red}{1}&\textcolor{red}{0}&0&0&\textcolor{red}{1}&\textcolor{red}{0}&0&0&\textcolor{red}{1}&\textcolor{red}{0}&0&0&\textcolor{red}{1}&\textcolor{red}{0}&0&0&\textcolor{red}{1}&\textcolor{red}{0}&0&0&\textcolor{red}{1}&\textcolor{red}{0}\\
0&0&\textcolor{red}{0}&\textcolor{red}{1}&0&0&\textcolor{red}{0}&\textcolor{red}{1}&0&0&\textcolor{red}{0}&\textcolor{red}{1}&0&0&\textcolor{red}{0}&\textcolor{red}{1}&0&0&\textcolor{red}{0}&\textcolor{red}{1}&0&0&\textcolor{red}{0}&\textcolor{red}{1}&0&0&\textcolor{red}{0}&\textcolor{red}{1}&0&0&\textcolor{red}{0}&\textcolor{red}{1}\\
1&0&0&0&1&0&0&0&1&0&0&0&1&0&0&0&1&0&0&0&1&0&0&0&1&0&0&0&1&0&0&0\\
0&1&0&0&0&1&0&0&0&1&0&0&0&1&0&0&0&1&0&0&0&1&0&0&0&1&0&0&0&1&0&0\\
0&0&\textcolor{red}{1}&\textcolor{red}{0}&0&0&\textcolor{red}{1}&\textcolor{red}{0}&0&0&\textcolor{red}{1}&\textcolor{red}{0}&0&0&\textcolor{red}{1}&\textcolor{red}{0}&0&0&\textcolor{red}{1}&\textcolor{red}{0}&0&0&\textcolor{red}{1}&\textcolor{red}{0}&0&0&\textcolor{red}{1}&\textcolor{red}{0}&0&0&\textcolor{red}{1}&\textcolor{red}{0}\\
0&0&\textcolor{red}{0}&\textcolor{red}{1}&0&0&\textcolor{red}{0}&\textcolor{red}{1}&0&0&\textcolor{red}{0}&\textcolor{red}{1}&0&0&\textcolor{red}{0}&\textcolor{red}{1}&0&0&\textcolor{red}{0}&\textcolor{red}{1}&0&0&\textcolor{red}{0}&\textcolor{red}{1}&0&0&\textcolor{red}{0}&\textcolor{red}{1}&0&0&\textcolor{red}{0}&\textcolor{red}{1}
\end{bmatrix}\]

\[
\setlength{\arraycolsep}{1pt}
H_X'=\begin{bmatrix}
1&0&0&0&1&0&0&0&1&0&0&0&1&0&0&0&1&0&0&0&1&0&0&0&1&0&0&0&1&0&0&0\\
0&1&0&0&0&1&0&0&0&1&0&0&0&1&0&0&0&1&0&0&0&1&0&0&0&1&0&0&0&1&0&0\\
0&0&\textcolor{red}{1}&\textcolor{red}{0}&0&0&\textcolor{red}{1}&\textcolor{red}{0}&0&0&\textcolor{red}{1}&\textcolor{red}{0}&0&0&\textcolor{red}{1}&\textcolor{red}{0}&0&0&\textcolor{red}{1}&\textcolor{red}{0}&0&0&\textcolor{red}{1}&\textcolor{red}{0}&0&0&\textcolor{red}{1}&\textcolor{red}{0}&0&0&\textcolor{red}{1}&\textcolor{red}{0}\\
0&0&\textcolor{red}{0}&\textcolor{red}{1}&0&0&\textcolor{red}{0}&\textcolor{red}{1}&0&0&\textcolor{red}{0}&\textcolor{red}{1}&0&0&\textcolor{red}{0}&\textcolor{red}{1}&0&0&\textcolor{red}{0}&\textcolor{red}{1}&0&0&\textcolor{red}{0}&\textcolor{red}{1}&0&0&\textcolor{red}{0}&\textcolor{red}{1}&0&0&\textcolor{red}{0}&\textcolor{red}{1}\\
1&0&0&0&1&0&0&0&1&0&0&0&1&0&0&0&1&0&0&0&1&0&0&0&1&0&0&0&1&0&0&0\\
0&1&0&0&0&1&0&0&0&1&0&0&0&1&0&0&0&1&0&0&0&1&0&0&0&1&0&0&0&1&0&0\\
0&0&\textcolor{red}{1}&\textcolor{red}{0}&0&0&\textcolor{red}{1}&\textcolor{red}{0}&0&0&\textcolor{red}{1}&\textcolor{red}{0}&0&0&\textcolor{red}{1}&\textcolor{red}{0}&0&0&\textcolor{red}{1}&\textcolor{red}{0}&0&0&\textcolor{red}{1}&\textcolor{red}{0}&0&0&\textcolor{red}{1}&\textcolor{red}{0}&0&0&\textcolor{red}{1}&\textcolor{red}{0}\\
0&0&\textcolor{red}{0}&\textcolor{red}{1}&0&0&\textcolor{red}{\mathbf{1}}&\textcolor{red}{1}&0&0&\textcolor{red}{0}&\textcolor{red}{1}&0&0&\textcolor{red}{0}&\textcolor{red}{1}&0&0&\textcolor{red}{0}&\textcolor{red}{1}&0&0&\textcolor{red}{0}&\textcolor{red}{\mathbf{0}}&0&0&\textcolor{red}{0}&\textcolor{red}{1}&0&0&\textcolor{red}{0}&\textcolor{red}{1}\\
1&0&0&0&1&0&0&0&1&0&0&0&1&0&0&0&1&0&0&0&1&0&0&0&1&0&0&0&1&0&0&0\\
0&1&0&0&0&1&0&0&0&1&0&0&0&1&0&0&0&1&0&0&0&1&0&0&0&1&0&0&0&1&0&0\\
0&0&\textcolor{red}{1}&\textcolor{red}{0}&0&0&\textcolor{red}{\mathbf{0}}&\textcolor{red}{0}&0&0&\textcolor{red}{1}&\textcolor{red}{0}&0&0&\textcolor{red}{1}&\textcolor{red}{0}&0&0&\textcolor{red}{1}&\textcolor{red}{0}&0&0&\textcolor{red}{1}&\textcolor{red}{\mathbf{1}}&0&0&\textcolor{red}{1}&\textcolor{red}{0}&0&0&\textcolor{red}{1}&\textcolor{red}{0}\\
0&0&\textcolor{red}{0}&\textcolor{red}{1}&0&0&\textcolor{red}{0}&\textcolor{red}{1}&0&0&\textcolor{red}{0}&\textcolor{red}{1}&0&0&\textcolor{red}{0}&\textcolor{red}{1}&0&0&\textcolor{red}{0}&\textcolor{red}{1}&0&0&\textcolor{red}{0}&\textcolor{red}{1}&0&0&\textcolor{red}{0}&\textcolor{red}{1}&0&0&\textcolor{red}{0}&\textcolor{red}{1}
\end{bmatrix}\]

\medskip
\noindent\textbf{Repair.}
We solve a compact ILP on the submatrix $\Delta_{I,J}$ to restore $H_X' H_Z^{\rm T}=0$ while preserving the row/column weights of $H_Z$.
In this instance, $|I|=6$, $|J|=16$, hence $v=|I||J|=96$ variables; the parity matrix has rank $31$, yielding $96-31=65$ degrees of freedom.
One feasible $\Delta$ that also satisfies the weight-balance constraints is shown below. 
\[
\setlength{\arraycolsep}{1pt}
\Delta=
\begin{bmatrix}
0&0&0&0&0&0&0&0&0&0&0&0&0&0&0&0&0&0&0&0&0&0&0&0&0&0&0&0&0&0&0&0\\
0&0&0&0&0&0&0&0&0&0&0&0&0&0&0&0&0&0&0&0&0&0&0&0&0&0&0&0&0&0&0&0\\
0&0&\textcolor{red}{0}&\textcolor{red}{0}&0&0&\textcolor{red}{1}&\textcolor{red}{1}&0&0&\textcolor{red}{1}&\textcolor{red}{0}&0&0&\textcolor{red}{0}&\textcolor{red}{1}&0&0&\textcolor{red}{0}&\textcolor{red}{0}&0&0&\textcolor{red}{0}&\textcolor{red}{0}&0&0&\textcolor{red}{0}&\textcolor{red}{0}&0&0&\textcolor{red}{0}&\textcolor{red}{0}\\
0&0&\textcolor{red}{0}&\textcolor{red}{1}&0&0&\textcolor{red}{1}&\textcolor{red}{1}&0&0&\textcolor{red}{1}&\textcolor{red}{0}&0&0&\textcolor{red}{0}&\textcolor{red}{0}&0&0&\textcolor{red}{0}&\textcolor{red}{0}&0&0&\textcolor{red}{0}&\textcolor{red}{0}&0&0&\textcolor{red}{0}&\textcolor{red}{0}&0&0&\textcolor{red}{0}&\textcolor{red}{0}\\
0&0&0&0&0&0&0&0&0&0&0&0&0&0&0&0&0&0&0&0&0&0&0&0&0&0&0&0&0&0&0&0\\
0&0&0&0&0&0&0&0&0&0&0&0&0&0&0&0&0&0&0&0&0&0&0&0&0&0&0&0&0&0&0&0\\
0&0&\textcolor{red}{0}&\textcolor{red}{1}&0&0&\textcolor{red}{1}&\textcolor{red}{0}&0&0&\textcolor{red}{0}&\textcolor{red}{0}&0&0&\textcolor{red}{1}&\textcolor{red}{0}&0&0&\textcolor{red}{0}&\textcolor{red}{1}&0&0&\textcolor{red}{0}&\textcolor{red}{0}&0&0&\textcolor{red}{0}&\textcolor{red}{0}&0&0&\textcolor{red}{0}&\textcolor{red}{0}\\
0&0&\textcolor{red}{1}&\textcolor{red}{0}&0&0&\textcolor{red}{1}&\textcolor{red}{0}&0&0&\textcolor{red}{0}&\textcolor{red}{0}&0&0&\textcolor{red}{0}&\textcolor{red}{1}&0&0&\textcolor{red}{0}&\textcolor{red}{1}&0&0&\textcolor{red}{0}&\textcolor{red}{0}&0&0&\textcolor{red}{0}&\textcolor{red}{0}&0&0&\textcolor{red}{0}&\textcolor{red}{0}\\
0&0&0&0&0&0&0&0&0&0&0&0&0&0&0&0&0&0&0&0&0&0&0&0&0&0&0&0&0&0&0&0\\
0&0&0&0&0&0&0&0&0&0&0&0&0&0&0&0&0&0&0&0&0&0&0&0&0&0&0&0&0&0&0&0\\
0&0&\textcolor{red}{1}&\textcolor{red}{1}&0&0&\textcolor{red}{1}&\textcolor{red}{1}&0&0&\textcolor{red}{0}&\textcolor{red}{0}&0&0&\textcolor{red}{0}&\textcolor{red}{0}&0&0&\textcolor{red}{0}&\textcolor{red}{0}&0&0&\textcolor{red}{0}&\textcolor{red}{0}&0&0&\textcolor{red}{0}&\textcolor{red}{0}&0&0&\textcolor{red}{0}&\textcolor{red}{0}\\
0&0&\textcolor{red}{0}&\textcolor{red}{1}&0&0&\textcolor{red}{1}&\textcolor{red}{1}&0&0&\textcolor{red}{0}&\textcolor{red}{0}&0&0&\textcolor{red}{1}&\textcolor{red}{0}&0&0&\textcolor{red}{0}&\textcolor{red}{0}&0&0&\textcolor{red}{0}&\textcolor{red}{0}&0&0&\textcolor{red}{0}&\textcolor{red}{0}&0&0&\textcolor{red}{0}&\textcolor{red}{0}
\end{bmatrix}
\]

\medskip
\noindent\textbf{Result.}
Applying $H_Z' = H_Z \oplus \Delta$ yields an $H_Z'$ that is orthogonal to $H_X'$ and preserves all row/column weights. 
\[
\setlength{\arraycolsep}{1pt}
H_Z'=
\begin{bmatrix}
1&0&0&0&1&0&0&0&1&0&0&0&1&0&0&0&1&0&0&0&1&0&0&0&1&0&0&0&1&0&0&0\\
0&1&0&0&0&1&0&0&0&1&0&0&0&1&0&0&0&1&0&0&0&1&0&0&0&1&0&0&0&1&0&0\\
0&0&\textcolor{red}{1}&\textcolor{red}{0}&0&0&\textcolor{red}{0}&\textcolor{red}{1}&0&0&\textcolor{red}{0}&\textcolor{red}{0}&0&0&\textcolor{red}{1}&\textcolor{red}{1}&0&0&\textcolor{red}{1}&\textcolor{red}{0}&0&0&\textcolor{red}{1}&\textcolor{red}{0}&0&0&\textcolor{red}{1}&\textcolor{red}{0}&0&0&\textcolor{red}{1}&\textcolor{red}{0}\\
0&0&\textcolor{red}{0}&\textcolor{red}{0}&0&0&\textcolor{red}{1}&\textcolor{red}{0}&0&0&\textcolor{red}{1}&\textcolor{red}{1}&0&0&\textcolor{red}{0}&\textcolor{red}{1}&0&0&\textcolor{red}{0}&\textcolor{red}{1}&0&0&\textcolor{red}{0}&\textcolor{red}{1}&0&0&\textcolor{red}{0}&\textcolor{red}{1}&0&0&\textcolor{red}{0}&\textcolor{red}{1}\\
1&0&0&0&1&0&0&0&1&0&0&0&1&0&0&0&1&0&0&0&1&0&0&0&1&0&0&0&1&0&0&0\\
0&1&0&0&0&1&0&0&0&1&0&0&0&1&0&0&0&1&0&0&0&1&0&0&0&1&0&0&0&1&0&0\\
0&0&\textcolor{red}{1}&\textcolor{red}{1}&0&0&\textcolor{red}{0}&\textcolor{red}{0}&0&0&\textcolor{red}{1}&\textcolor{red}{0}&0&0&\textcolor{red}{0}&\textcolor{red}{0}&0&0&\textcolor{red}{1}&\textcolor{red}{1}&0&0&\textcolor{red}{1}&\textcolor{red}{0}&0&0&\textcolor{red}{1}&\textcolor{red}{0}&0&0&\textcolor{red}{1}&\textcolor{red}{0}\\
0&0&\textcolor{red}{1}&\textcolor{red}{1}&0&0&\textcolor{red}{1}&\textcolor{red}{1}&0&0&\textcolor{red}{0}&\textcolor{red}{1}&0&0&\textcolor{red}{0}&\textcolor{red}{0}&0&0&\textcolor{red}{0}&\textcolor{red}{0}&0&0&\textcolor{red}{0}&\textcolor{red}{1}&0&0&\textcolor{red}{0}&\textcolor{red}{1}&0&0&\textcolor{red}{0}&\textcolor{red}{1}\\
1&0&0&0&1&0&0&0&1&0&0&0&1&0&0&0&1&0&0&0&1&0&0&0&1&0&0&0&1&0&0&0\\
0&1&0&0&0&1&0&0&0&1&0&0&0&1&0&0&0&1&0&0&0&1&0&0&0&1&0&0&0&1&0&0\\
0&0&\textcolor{red}{0}&\textcolor{red}{1}&0&0&\textcolor{red}{0}&\textcolor{red}{1}&0&0&\textcolor{red}{1}&\textcolor{red}{0}&0&0&\textcolor{red}{1}&\textcolor{red}{0}&0&0&\textcolor{red}{1}&\textcolor{red}{0}&0&0&\textcolor{red}{1}&\textcolor{red}{0}&0&0&\textcolor{red}{1}&\textcolor{red}{0}&0&0&\textcolor{red}{1}&\textcolor{red}{0}\\
0&0&\textcolor{red}{0}&\textcolor{red}{0}&0&0&\textcolor{red}{1}&\textcolor{red}{0}&0&0&\textcolor{red}{0}&\textcolor{red}{1}&0&0&\textcolor{red}{1}&\textcolor{red}{1}&0&0&\textcolor{red}{0}&\textcolor{red}{1}&0&0&\textcolor{red}{0}&\textcolor{red}{1}&0&0&\textcolor{red}{0}&\textcolor{red}{1}&0&0&\textcolor{red}{0}&\textcolor{red}{1}\\
\end{bmatrix}\]
\medskip
\noindent\textbf{Remark.}
In this workflow the ILP dimension $v=|I||J|$ and the number of parity constraints depend only on the maximum column/row weights $(d_c,d_r)$ and are independent of the global block size $P$, enabling scalability.

After several hundred random switch–repair iterations, we obtained the following matrices, which form a well-randomized orthogonal pair.
\begin{align}
 \setlength{\arraycolsep}{1pt}
H_X=
\begin{bmatrix}
 0&0&0&0&0&1&0&0&0&0&0&0&0&1&0&1&0&0&0&0&1&0&0&0&1&1&0&0&0&0&1&1\\
 0&0&0&0&0&0&1&0&0&0&1&0&1&1&0&0&1&0&0&1&0&0&0&0&0&0&0&1&1&0&0&0\\
 1&0&0&0&0&0&0&1&0&1&1&0&0&0&0&0&1&0&0&0&0&1&0&0&1&0&1&0&0&0&0&0\\
 1&0&0&1&0&1&0&0&0&0&0&1&0&0&0&0&0&0&0&0&1&0&0&1&1&0&1&0&0&0&0&0\\
 0&0&1&0&0&0&0&0&1&0&1&0&0&0&1&0&0&0&1&0&0&1&0&0&0&0&0&0&0&1&0&1\\
 1&1&0&0&1&0&1&1&0&1&0&0&0&0&0&0&0&0&0&0&0&0&0&1&0&1&0&0&0&0&0&0\\
 0&0&0&0&0&0&0&0&0&0&0&0&0&0&1&0&0&1&1&1&0&0&1&1&0&0&0&0&0&0&1&1\\
 0&0&1&0&0&0&1&1&0&0&0&1&1&0&0&0&1&0&0&1&0&0&1&0&0&0&0&0&0&0&0&0\\
 0&1&0&1&0&1&0&0&0&0&0&0&0&0&0&0&0&0&0&0&1&0&0&0&0&1&0&1&1&1&0&0\\
 0&1&0&0&1&0&0&0&1&0&0&0&0&0&1&0&0&1&0&0&0&0&1&0&0&0&0&0&1&0&1&0\\
 0&0&1&0&0&0&0&0&0&1&0&1&0&0&0&1&0&0&1&0&0&0&0&0&0&0&1&1&0&1&0&0\\
 0&0&0&1&1&0&0&0&1&0&0&0&1&1&0&1&0&1&0&0&0&1&0&0&0&0&0&0&0&0&0&0
\end{bmatrix}
\end{align}
\begin{align}
 H_Z=
 \setlength{\arraycolsep}{1pt}
\begin{bmatrix}
 0&1&0&0&0&1&1&0&0&0&0&1&0&0&0&0&0&0&1&0&0&0&0&0&0&0&0&1&0&1&1&0\\
 0&0&0&0&0&0&1&1&1&0&0&0&0&0&1&0&0&1&0&0&1&0&0&0&1&0&0&0&1&0&0&0\\
 1&0&0&0&0&1&0&0&1&0&1&0&1&0&0&0&0&0&0&0&1&0&1&1&0&0&0&0&0&0&0&0\\
 1&1&0&0&0&0&0&0&0&0&0&0&0&0&1&0&1&0&0&1&0&0&0&1&0&1&0&0&0&0&0&1\\
 0&0&1&1&0&1&0&0&0&0&0&1&0&0&0&0&0&0&1&0&0&1&0&0&0&0&1&0&0&0&0&1\\
 0&0&1&0&1&0&0&0&0&0&1&1&0&0&0&1&0&0&0&0&0&0&0&0&0&1&1&0&1&0&0&0\\
 0&0&0&0&1&0&0&0&1&1&0&0&1&0&1&1&1&0&0&0&0&0&0&0&0&0&0&0&0&0&1&0\\
 0&0&0&1&0&0&0&1&0&1&0&0&0&0&0&0&0&0&0&0&0&1&1&0&1&0&0&0&0&1&1&0\\
 1&0&0&0&0&0&1&0&0&0&0&0&0&1&0&0&1&0&0&0&0&1&0&0&1&0&0&1&0&1&0&0\\
 0&0&1&0&0&0&0&0&0&1&1&0&0&1&0&0&0&1&0&1&0&0&0&0&0&1&0&0&1&0&0&0\\
 0&1&0&1&0&0&0&1&0&0&0&0&0&1&0&1&0&1&0&1&0&0&0&0&0&0&1&0&0&0&0&0\\
 0&0&0&0&1&0&0&0&0&0&0&0&1&0&0&0&0&0&1&0&1&0&1&1&0&0&0&1&0&0&0&1
\end{bmatrix}
\end{align}

For this final $12\times32$ pair, direct binary row reduction gives
\[
\operatorname{rank}H_X=\operatorname{rank}H_Z=12,
\qquad H_XH_Z^{\mathsf T}=0,
\qquad k=32-12-12=8.
\]
Exhaustive enumeration of the two $20$-dimensional kernels gives the exact CSS distances
\[
d_X=4,\qquad d_Z=2,
\]
and hence the final pair defines a $[[32,8,2]]$ code. With zero-based column indices, an $X$-type minimum-weight representative is supported on $\{4,5,11,12\}$, and a $Z$-type minimum-weight representative is supported on $\{5,20\}$. The enumeration checks every nonzero vector in $\ker H_Z$ modulo $\operatorname{row}(H_X)$ and every nonzero vector in $\ker H_X$ modulo $\operatorname{row}(H_Z)$.

\section{Conclusion and Future Work}

We have presented a randomized construction method for orthogonal sparse matrix pairs that preserves matrix weight distributions exactly while introducing local randomness.
This property is crucial for exploring the ensemble of random quantum LDPC codes consistent with a given structural design.
The proposed algorithm performs local \(2\times2\) switches followed by ILP-based orthogonality repair, with computational complexity depending solely on the maximum row and column weights.

Since the randomized matrices maintain their degree distributions, the decoding performance is expected to be well described by density-evolution analysis.
Furthermore, the randomization has the potential to lower the error floor by eliminating small trapping sets while maintaining scalability.

In future work, we plan to extend this framework to general stabilizer codes and to perform decoding experiments using both binary and non-binary belief-propagation algorithms.
Since the proposed local repair formulation can flexibly incorporate additional constraints into the ILP, it can also be adapted to codes with structured connectivity such as spatially coupled \cite{CSS_SC_ISIT}or multi-edge type LDPC codes \cite{Richardson2002}.
This enables the integration of classical near-capacity code design techniques into quantum LDPC constructions.
Furthermore, designing matrices that avoid short cycles and increase the minimum distance will be an important direction.
We believe that these extended formulations can be implemented without difficulty thanks to the flexibility of the ILP-based approach.
\bibliographystyle{IEEEtran}
\bibliography{IEEEabrv,00kasai} 
\end{document}